%% file: jrssa_paper.tex
\documentclass[webpdf,modern,medium]{oup-authoring-template}
\usepackage{bm}
\usepackage{float}
\usepackage{booktabs}
\usepackage{placeins}

\onecolumn 

\graphicspath{{Fig/}}

\theoremstyle{thmstyleone}%
\newtheorem{theorem}{Theorem}
\theoremstyle{thmstyletwo}%
\newtheorem{remark}{Remark}%
\newtheorem{corollary}{Corollary}

\newtheorem{lemma}{Lemma}
\theoremstyle{thmstylethree}%

\newcommand{\bx}{\mathbf{x}}

\renewcommand{\bf}[1]{\mathbf{#1}}
\DeclareMathOperator*{\logistic}{logistic}
\DeclareMathOperator*{\logit}{logit}
\DeclareMathOperator*{\argmin}{arg\;min}
\DeclareMathOperator*{\argmax}{arg\;max}

\DeclareMathOperator*{\Var}{Var}

\begin{document}

\journaltitle{Under revision at: The Journal of the Royal Statistical Society, Series A (Statistics in Society)}
\DOI{DOI added during production}
\copyrightyear{2026}
\pubyear{YEAR}
\vol{XX}
\issue{x}
\access{Published: Date added during production}
\appnotes{Paper}

\firstpage{1}


\title[Data integration under MAR]{Data integration using propensity score weighting under missing at random}

\author[1,2$\ast$]{Caleb Leedy\ORCID{0009-0001-5809-6805}}
\author[1]{Jae Kwang Kim}

\address[1]{\orgdiv{Department of Statistics}, \orgname{Iowa State University}, \orgaddress{\street{2434 Osborn Drive}, \postcode{50010}, \state{IA}, \country{USA}}}
\address[2]{\orgdiv{Department of Mathematics and Statistics}, \orgname{University of Ottawa}, \orgaddress{\street{150 Louis-Pasteur Pvt}, \postcode{K1N 6N5}, \state{ON}, \country{Canada}}}

\corresp[$\ast$]{Corresponding author. \href{cleedy@uottawa.ca}{cleedy@uottawa.ca}}

\received{Date}{0}{Year}
\revised{Date}{0}{Year}
\accepted{Date}{0}{Year}



\abstract{ 
We study data integration for the setting in which a probability sample and a
nonprobability sample are both available from a common target population, with
auxiliary variables observed in both samples and the study variable observed
only in the nonprobability sample. Under a parametric propensity score model
for the nonprobability sample inclusion mechanism, we develop a conditional
maximum likelihood estimator of the inclusion probabilities and construct an
inverse probability weighted (IPW) estimator of the population mean. We
establish consistency and asymptotic normality of the proposed IPW estimator
under suitable regularity conditions. To guard against misspecification of the propensity score
model, we further incorporate an outcome regression model through debiased
generalized entropy calibration (GEC) 
yielding a doubly robust estimator that remains consistent when the
propensity score model or the outcome regression model is correctly specified.
We also develop variance estimation procedures for both estimators. Simulation
studies illustrate the finite-sample performance of the proposed methods, and
we apply them to data from the Culture and Community in a Time of Crisis
survey, where the nonprobability sample substantially over-represents certain
demographic groups relative to the U.S.\ adult population.
}

\keywords{
survey sampling,
data integration,
doubly robust estimation,
missing at random,
nonprobability samples,
propensity score weighting
}





\maketitle

\section{Introduction}

Probability sampling is the gold standard for producing representative data and
supporting design-consistent statistical inference \citep{fuller2009sampling}.
In many modern applications, however, probability samples are expensive to
collect and often limited in size. These limitations led to early work
combining probability samples with other nonprobability samples 
\citep{rivers2007sampling, schenker2007combining, elliot2009combining, dong2014combining}.
These nonprobability samples such as
online opt-in panels, administrative platforms, and convenience samples can be
obtained more easily and may contain large numbers of observations. The main
drawback is that nonprobability samples are generally affected by unknown
selection bias \citep{beaumont2021pitfalls} 
and therefore cannot, by themselves, support valid population inference.

When a probability sample and a nonprobability sample are both available from
the same target population, data integration methods provide a way to combine
the strengths of the two sources. The probability sample contributes
representativeness and design information, while the nonprobability sample
contributes outcome information and, often, a larger effective sample size. 
\cite{yang2020statistical}, \cite{wu2022statistical} and \cite{paddock2025} 
provide comprehensive overview of the literature in data integration. 

In this paper, we study the setting in which auxiliary variables 
$X$ are observed in both samples, while the study variable $Y$ is observed only
in the nonprobability sample. This is one of the central settings in survey data
integration, because inference on the population mean of 
$Y$ requires correcting for the unknown selection mechanism of the nonprobability sample. 

Two broad strategies have been developed for this problem \citep{rao2021making}.
The first is outcome
modeling, often implemented through mass imputation, where one models the
conditional distribution of $Y$ given $X$ using the nonprobability sample and
then predicts the missing values of 
$Y$ in the probability sample. See, for example, \cite{tsung2018model},
\cite{chen2019calibrating}, \cite{kim2021combining} and
\cite{chen2022nonparametric}. The second is propensity score adjustment 
\citep{valliant2011estimating, elliott2017inference}, where
one models the inclusion probability for the nonprobability sample and then uses
inverse probability weighting to remove selection bias. \cite{chen2020doubly}
developed a pseudo-maximum likelihood approach for this problem and established
formal large-sample properties. More recently, \cite{kim2024comments}  and
\cite{beaumont2024} considered conditional likelihood formulations that use the
joint structure of the two samples more efficiently. 

A natural goal in this context is to combine the advantages of the two
approaches. Propensity score weighting is attractive when the selection
mechanism can be modeled well, but it can be sensitive to misspecification of
the inclusion model. Outcome-model-based methods can be efficient, but they rely
on correct specification of the regression model for 
$Y \mid X$. This motivates doubly robust procedures, which remain consistent
when either the sampling score model or the outcome regression model is
correctly specified. Doubly robust ideas are well established in missing data
analysis and survey sampling; see 
\cite{robins1994estimation}, \cite{kimhaziza12},  and
\cite{kang2007demystifying}. The challenge in the present data integration
problem is to develop such estimators in a way that is compatible with the
conditional likelihood approach for the nonprobability sampling mechanism and
that also admits workable variance estimation.

The starting point of this paper is the conditional likelihood formulation of
\cite{kim2024comments}, which views the observed samples through a two-phase
structure. This perspective leads to an estimator of the nonprobability sampling
score that differs from the pseudo-likelihood approach of \cite{chen2020doubly} and
can make more efficient use of the available information. Building on this idea,
we first develop an inverse probability weighted estimator for the population
mean based on the conditional maximum likelihood estimator of the sampling
score. We derive an asymptotic linear representation and a design-based variance
estimator under the proposed framework. We then extend this approach by
incorporating generalized entropy calibration, following
\cite{kwon2024debiased}, to construct a calibration estimator that uses both
covariate balancing and the estimated sampling score. This yields a doubly
robust estimation strategy that is consistent when either the sampling score
model or the outcome regression model is correctly specified, under the
regularity conditions stated in the paper.

The main contributions of the paper are therefore threefold. First, we provide a
conditional-likelihood-based inverse probability weighting estimator for data
integration when 
Y is observed only in the nonprobability sample. Second, we combine this
conditional likelihood framework with generalized entropy calibration to obtain
a doubly robust estimator that incorporates both sampling-score information and
outcome-model information. Third, we develop corresponding variance estimators
and study their empirical performance through simulations and an application to
the Culture and Community in a Time of Crisis survey. Relative to the
pseudo-likelihood approach of \cite{chen2020doubly}, the proposed
conditional-likelihood formulation provides an alternative way to estimate the
sampling score and to construct inverse probability weighted and doubly robust
estimators for the population mean. Our focus is on consistency, asymptotic
normality, and double robustness under potential misspecification of the working models.

The remainder of the paper is organized as follows. Section 2 introduces the
basic setup and reviews existing propensity-score-based estimation for data
integration. Section 3 develops the conditional-likelihood-based inverse
probability weighting estimator and its variance estimator. Section 4 presents
the generalized entropy calibration extension and studies its doubly robust
properties. Section 5 reports simulation results, and Section 6 presents the
application to the Culture and Community in a Time of Crisis survey. Section 7
concludes with a discussion of limitations and future work.

\section{Basic Setup}

We adopt the data integration framework introduced by
\cite{yang2020statistical}, which considers the setting in which two independent
samples are drawn from a common finite population. The first sample, denoted as
Sample A, is a probability sample for which covariates $\bf{X} \in \mathbb{R}^m$
are observed. The second sample, Sample B, is a non-probability sample for which
both $\bf{X}$ and the outcome variable $Y$ are observed. Table~\ref{tab:setup}
summarizes the data structure in the two samples.

\begin{table}[ht!]
  \centering
  \begin{tabular}{lccr}
    \toprule
    Sample & $X$ & $Y$ & Type \\
    \midrule
    A & \checkmark &  & Probability \\
    B & \checkmark & \checkmark & Nonprobability \\
    \bottomrule
  \end{tabular}
  \caption{Data Integration Problem Setup.}
  \label{tab:setup}
\end{table}

Because Sample B is drawn through an uncontrolled, non-probabilistic mechanism,
it is not representative of the target population. Consequently, naive inference
based solely on Sample B is generally biased. Our goal is to estimate the
population mean $\theta = N^{-1} \sum_{i \in U} y_i$ by combining information
from both samples.

Let $U = \{1, \dots, N\}$ denote the index set of the target finite population.
Let $\bx_i$ and $y_i$ be the vector of auxiliary variables and the study
variable of interest, respectively, associated with unit $i \in U$. Let
$\delta^{(A)}_i = 1$ if unit $i$ is included in Sample A and $0$ otherwise, and
define $\delta^{(B)}_i$ analogously for Sample B. We assume that for units in
Sample A, the inclusion probabilities $\pi^{(A)}_i = \Pr(\delta^{(A)}_i = 1)$
are known from the sampling design. Let $n_A$ and $n_B$ denote the sample sizes
of Sample A and Sample B, respectively.

To accomplish this goal, we model the probability of inclusion in Sample B using
a parametric propensity score model. Specifically, we assume that the sampling
mechanism can be expressed as $\pi^{(B)}_i = \pi^{(B)}(x_i; \bm \phi)$ for some
unknown parameter $\bm \phi \in \mathbb{R}^m$. Under this model, we apply
inverse probability weighting (IPW) to adjust for the selection bias inherent in
Sample B.

Before developing our approach, we first review the pseudo-maximum likelihood
framework of \citet{chen2020doubly} (CLW), which serves as the primary point of
comparison. They assume a parametric form for $\pi^{(B)}_i(\bm \phi)$ and
consider the log-likelihood function:
\begin{align}\label{eq:cheneq}
  \ell(\bm \phi)
  &= \sum_{i = 1}^N \left\{\delta_i^{(B)} \log(\pi_i^{(B)}(\bm \phi)) +
    (1 - \delta_i^{(B)}) \log(1 - \pi_i^{(B)}(\bm \phi))\right\} \\
  &= \sum_{i \in B} \log \left(\frac{\pi_i^{(B)}(\bm \phi)}{1 - \pi_i^{(B)}(\bm \phi)}\right) +
  \sum_{i \in U} \log(1 - \pi_i^{(B)}(\bm \phi)). \label{eq:cheneq2}
\end{align}
Since the full population is not observed, the second term of
(\ref{eq:cheneq2}) cannot be computed directly. To overcome this, CLW estimate
it using Horvitz--Thompson estimation from Sample A, leading to the pseudo
maximum likelihood estimator (PMLE):
\begin{equation}
  \hat{\bm \phi} = \arg\max_{\bm \phi} \left\{ \sum_{i \in B} \log \left(
  \frac{\pi^{(B)}_i(\bm \phi)}{1 - \pi^{(B)}_i(\bm \phi)} \right) + \sum_{i \in
A} \frac{1}{\pi^{(A)}_i} \log (1 - \pi^{(B)}_i(\bm \phi)) \right\}.
\label{eq:chenee}
\end{equation}
While the PMLE is consistent, it is not necessarily efficient because it does
not optimally use information from both samples to approximate the
log-likelihood. \citet{kim2024comments} and \cite{beaumont2024} independently
propose an improvement by developing a conditional maximum likelihood estimator
(CMLE) of ${\bm \phi}$, which leads to more efficient estimation of the
propensity score model and enables the derivation of valid variance estimators.
Details of this framework and the associated variance estimation techniques are
developed in Section~\ref{sec:ipw}.

\section{Inverse Probability Weighting Estimation}
\label{sec:ipw}

We now develop the inverse probability weighting (IPW) estimator for the
population mean $\theta = E[Y]$, building on the conditional maximum likelihood
framework introduced by \citet{kim2024comments}.

Let $S_1 = A \cup B$ denote the union of the probability and non-probability
samples. In this framework, Sample $S_2 = B$ is viewed as a second-phase sample
drawn from $S_1$, allowing us to construct a conditional likelihood based on
observed inclusion indicators. Define the first- and second-phase inclusion
probabilities as follows:
\[
\pi^{(1)}_i = \Pr(i \in S_1), \quad \pi^{(2|1)}_i = \Pr(i \in S_2 \mid i \in S_1).
\]
By construction, the first-phase inclusion probability satisfies
\begin{equation}\label{eq:pi1}
  \pi_i^{(1)}(\bm \phi) = \pi_i^{(A)} + \pi_i^{(B)}(\bm \phi) - \pi_i^{(A)}
  \pi_i^{(B)}(\bm \phi),
\end{equation}
and the conditional inclusion probability becomes
\begin{equation}\label{eq:pi21}
  \pi_i^{(2|1)}(\bm \phi) = \frac{\pi_i^{(B)}(\bm \phi)}{\pi_i^{(1)}(\bm \phi)} = 
  \frac{\pi_i^{(B)}(\bm \phi)}{\pi_i^{(A)} + \pi_i^{(B)}(\bm \phi) - \pi_i^{(A)}\pi_i^{(B)}(\bm \phi)}.
\end{equation}

To estimate the parameter $\bm \phi$, we maximize the conditional log-likelihood
based on Sample $S_1$:
\begin{align}\label{eq:phihat}
  \hat{\bm \phi} 
  &= 
  \argmax_{\bm \phi}
    \sum_{i \in S_1} \{\delta_i^{(B)} \log \pi_i^{(2|1)}(\bm \phi) + (1 - \delta_i^{(B)}) \log (1 - \pi_i^{(2|1)}(\bm \phi))\}.
\end{align}
Once $\hat{\bm \phi}$ is obtained from (\ref{eq:phihat}), we estimate $\theta$
using the IPW estimator:
\begin{equation}\label{eq:ipwee}
\hat \theta_{\rm{IPW}} = 
\frac{\sum_{i \in B} y_i / \hat \pi^{(B)}_i}{\sum_{i \in B} 1 / \hat \pi^{(B)}_i}, 
\end{equation}
where $\hat \pi_i^{(B)} = \pi^{(B)}(\hat{\bm \phi}) = \pi^{(B)}(\bf x_i, \hat{\bm \phi})$.
The IPW estimator in (\ref{eq:ipwee}) can equivalently be obtained as the
solution to the system of estimating equations:
\begin{align}
  U_1(\theta, \bm \phi) &:= \frac{1}{N}  \sum_{i \in B} \frac{y_i - \theta}{\pi_i^{(B)}(\bm \phi)} = 0, \label{eq:eet1u1}\\
  U_2(\bm \phi) &:= \frac{1}{n_B} \sum_{i \in S_1} 
  (\delta_i^{(B)} - \pi_i^{(2|1)}(\bm \phi)) {\bf h}_i(\bm \phi) = \bf 0, \label{eq:eet1u2}
\end{align}
where $\bm{h}_i(\bm \phi) = \frac{\partial}{\partial \bm \phi} \text{logit}(\pi^{(2|1)}_i(\bm \phi))$.
As shown in the proof of Theorem~\ref{thm:1} in the online supplementary material, 
$\frac{\partial}{\partial \bm \phi} \text{logit}(\pi^{(2|1)}_i(\bm \phi)) = 
\frac{\partial}{\partial \bm \phi} \logit(\pi_i^{(B)}(\bm \phi))$,
so that ${\bf h}_i(\bm \phi)$ depends only on the posited propensity score model
$\pi_i^{(B)}(\bm \phi)$ and not on $\pi_i^{(A)}$.
Under the regularity conditions stated in the supplementary material, Theorem~\ref{thm:1}
provides a linearization of these estimating equations and establishes the
asymptotic structure of $\hat{\theta}_{\mathrm{IPW}}$. Let $\bm \phi^*$ denote
the probability limit of $\hat{\bm \phi}$, and define $\bf{h}_i(\bm \phi^*) =
h_i(\bm \phi) \big|_{\phi = \phi^*}$.

\begin{theorem}
\label{thm:1}
Let $n=\min(n_A,n_B)$. Under the regularity conditions stated in the supplementary material,
\begin{equation}
    \sqrt{n}\,(\hat{\theta}_{\emph{IPW}}-\theta)
    = -\frac{\sqrt{n}}{N}\,
    U_\ell(\theta,\phi^*,\gamma^*(\theta)) + o_p(1),
\end{equation}
where
\begin{align}
    U_\ell(\theta,\phi^*,\gamma^*)
    &:= U^{(A)}(\phi^*,\gamma^*) + U^{(B)}(\theta,\phi^*,\gamma^*), \\[4pt]
U^{(A)}(\phi^*,\gamma^*)
    &= \sum_{i\in A} \pi_i^{(2|1)}(\phi^*)\, h_i(\phi^*)^\top \gamma^*, \label{eq:UA_revised} \\[4pt] 
    U^{(B)}(\theta,\phi^*,\gamma^*)
    &= \sum_{i\in B}\frac{1}{\pi_i^{(B)}(\phi^*)}
       \left\{
       y_i-\theta
       -\pi_i^{(B)}(\phi^*)\bigl(1-\pi_i^{(2|1)}(\phi^*)\bigr)
       h_i(\phi^*)^\top\gamma^*
       \right\}, \label{eq:UB_revised}
\end{align} 
and $\gamma^*\equiv \gamma^*(\theta)$ is given by
\begin{equation}
    \gamma^*
    =
    \left(
      E\left[
      \sum_{i\in B}
      \pi_i^{(2|1)}(\phi^*)
      \bigl(1-\pi_i^{(2|1)}(\phi^*)\bigr)
      h_i(\phi^*)^{\otimes 2}
      \right]
    \right)^{-1}
    E\left[
      \sum_{i\in B}
      \frac{(y_i-\theta)\bigl(1-\pi_i^{(B)}(\phi^*)\bigr)}
           {\pi_i^{(B)}(\phi^*)}
      h_i(\phi^*)
    \right],
\end{equation}
with $h_i(\phi)^{\otimes 2}=h_i(\phi)h_i(\phi)^\top$.
\end{theorem}

Theorem~\ref{thm:1} gives an asymptotic linear representation of
$\hat{\theta}_{\mathrm{IPW}}$ through the decomposition
$U_\ell=U^{(A)}+U^{(B)}$. The term $U^{(A)}$ captures the contribution of
estimating the sampling score model through Sample $A$, while $U^{(B)}$
captures the variability of the inverse probability weighted estimating equation
based on Sample $B$. Since $\hat{\theta}_{\mathrm{IPW}}$ depends on both samples
through $(\hat{\theta}_{\mathrm{IPW}},\hat{\phi})$, we study the estimator under
the conventional two-sample asymptotic regime with
$n=\min(n_A,n_B)\to\infty$. Accordingly, the stochastic order of
$\hat{\theta}_{\mathrm{IPW}}-\theta$ is governed by the smaller of the two
sample sizes, and Theorem~\ref{thm:1} is stated at the root-$n$ scale.

\begin{remark}
\label{rem:overlap}
Let $n=\min(n_A,n_B)$. In the proof of Theorem~\ref{thm:1}, the estimating
equations are initially written as sums over $S_1=A\cup B$. Since a unit may
belong to both samples, these sums can be decomposed into contributions from
$A\setminus B$, $B\setminus A$, and $A\cap B$. The terms corresponding to
$A\cap B$ do not appear explicitly in the final linear representation because
their total contribution is asymptotically negligible under a mild overlap
condition.

Specifically, suppose that the contribution of a single overlapped unit to the
linearized estimating equation is uniformly $O_p(n^{-1})$. Then the total
contribution of the overlap is of order
\[
\sum_{i\in A\cap B} O_p(n^{-1})
=
O_p\!\left(\frac{|A\cap B|}{n}\right).
\]
Hence, if
\[
|A\cap B| = o_p(\sqrt{n}),
\]
then
\[
O_p\!\left(\frac{|A\cap B|}{n}\right)=o_p(n^{-1/2}),
\]
and the overlap terms are negligible at the root-$n$ scale. Therefore, the
linearization in Theorem~\ref{thm:1} may be written using only the separate
samplewise contributions $U^{(A)}$ and $U^{(B)}$, with the overlap absorbed
into the remainder term $o_p(n^{-1/2})$.
\end{remark}

The linearization in Theorem~\ref{thm:1}, together with the independence of
$U^{(A)}$ and $U^{(B)}$ yields the
following asymptotic normality result.

\begin{corollary}[Asymptotic normality of $\hat{\theta}_{\text{IPW}}$]
\label{cor:ipw_normality}
Under the conditions of Theorem~\ref{thm:1}, $\hat{\theta}_{\emph{IPW}}$ is
asymptotically unbiased for $\theta$, and
\begin{equation}
    \sqrt{n}\,(\hat{\theta}_{\emph{IPW}}-\theta)
    \xrightarrow{d} N(0,V_{\emph{IPW}}),
\end{equation}
where
\begin{equation}
    V_{\emph{IPW}}
    =
    \lim_{n\to\infty}
    \frac{n}{N^2}
    \left[
      \Var\!\left(U^{(A)}(\phi^*,\gamma^*)\right)
      +
      \Var\!\left(U^{(B)}(\theta,\phi^*,\gamma^*)\right)
    \right] 
\end{equation}
with
\begin{eqnarray}
\emph{Var}\!\left(U^{(A)}\right)
    &=&  \sum_{i \in U}\sum_{j \in U}
        \Delta^{(A)}_{ij}
        \left(\pi^{(2|1)}_i h_i(\phi^*)^\top \gamma^*\right)
        \left(\pi^{(2|1)}_j h_j(\phi^*)^\top \gamma^*\right), 
      \label{eq:varUA}  \\
        \emph{Var}\!\left(U^{(B)}\right)
    &=&  \sum_{i \in U}
        \frac{1 - \pi^{(B)}_i(\phi^*)}{\pi^{(B)}_i(\phi^*)}
        \left(e^{(B)}_i\right)^2,
        \label{eq:varUB}
\end{eqnarray}
where $\Delta^{(A)}_{ij} = \pi^{(A)}_{ij} - \pi^{(A)}_i \pi^{(A)}_j$ and
\begin{equation}
    e^{(B)}_i
    = y_i - \theta
      + \pi^{(B)}_i(\phi^*)\!\left(1 - \pi^{(2|1)}_i(\phi^*)\right)
        h_i(\phi^*)^\top \gamma^*.
        \label{eq:residual}
\end{equation}
\end{corollary}

Corollary~\ref{cor:ipw_normality} completes the asymptotic theory for
$\hat{\theta}_{\emph{IPW}}$. Asymptotic unbiasedness follows from the zero-mean
property of $U_\ell$: the contribution $U^{(B)}$ is unbiased by the
Horvitz--Thompson argument, and $U^{(A)}$ is unbiased because
$E[\delta^{(B)}_i \mid i \in S_1] = \pi^{(2|1)}_i$ by construction of the
conditional likelihood. The asymptotic variance $V_{\emph{IPW}}$ decomposes
additively into the two independent components identified in Theorem~\ref{thm:1}.
The term $\mathrm{Var}(U^{(A)})$ in~(\ref{eq:varUA}) reflects the uncertainty
from estimating $\phi$ using Sample $A$, which is of smaller order than the
dominant term $\mathrm{Var}(U^{(B)})$ in~(\ref{eq:varUB}) when $n_A/n_B \to 0$.
The residual $e^{(B)}_i$ in~(\ref{eq:residual}) equals $y_i - \theta$ when
$\phi$ is known, and acquires the correction term
$\pi^{(B)}_i(1 - \pi^{(2|1)}_i)h_i^\top\gamma^*$ to account for the estimation
of $\phi$.

We now construct consistent estimators of $V_{\emph{IPW}}$ by replacing the
unknown quantities $\phi^*$, $\gamma^*$, and $\theta$ with their sample
analogs. We estimate $\gamma^*$ by substituting $\hat{\phi}$ and
$\hat{\theta}_{\emph{IPW}}$ into the expression derived in the proof of
Theorem~\ref{thm:1}:
\begin{equation}
    \hat{\gamma}
    = \left(
        \sum_{i \in B}
        \hat{\pi}^{(2|1)}_i
        \bigl(1 - \hat{\pi}^{(2|1)}_i\bigr)
        h_i(\hat{\phi})^{\otimes 2}
      \right)^{\!-1}
      \sum_{i \in B}
      \frac{1 - \hat{\pi}^{(B)}_i}{\hat{\pi}^{(B)}_i}
      \bigl(y_i - \hat{\theta}_{\emph{IPW}}\bigr)
      h_i(\hat{\phi}).
      \label{eq:gamma_hat}
\end{equation}
The estimated variance contributions from Samples $A$ and $B$ are then
\begin{align}
    \widehat{\mathrm{Var}}\!\left(U^{(A)}\right)
    &= \hat{N}^{-2}
       \sum_{i \in A}\sum_{j \in A}
       \frac{\Delta^{(A)}_{ij}}{\pi^{(A)}_{ij}}
       \bigl(\hat{\pi}^{(2|1)}_i h_i(\hat{\phi})^\top \hat{\gamma}\bigr)
       \bigl(\hat{\pi}^{(2|1)}_j h_j(\hat{\phi})^\top \hat{\gamma}\bigr),
       \label{eq:varA_hat} \\[6pt]
    \widehat{\mathrm{Var}}\!\left(U^{(B)}\right)
    &= \hat{N}^{-2}
       \sum_{i \in B}
       \frac{1 - \hat{\pi}^{(B)}_i}{\bigl(\hat{\pi}^{(B)}_i\bigr)^2}
       \bigl(\hat{e}^{(B)}_i\bigr)^2,
       \label{eq:varB_hat}
\end{align}
where $\Delta^{(A)}_{ij} = \pi^{(A)}_{ij} - \pi^{(A)}_i\pi^{(A)}_j$,
$\hat{N} = \sum_{i \in B} 1/\hat{\pi}^{(B)}_i$, and
\begin{equation}
    \hat{e}^{(B)}_i
    = y_i - \hat{\theta}_{\emph{IPW}}
      + \hat{\pi}^{(B)}_i
        \bigl(1 - \hat{\pi}^{(2|1)}_i\bigr)
        h_i(\hat{\phi})^\top \hat{\gamma}.
    \label{eq:eB_hat}
\end{equation}
The total estimated variance is
\begin{equation}
    \widehat{\mathrm{Var}}(\hat{\theta}_{\emph{IPW}})
    = \widehat{\mathrm{Var}}\!\left(U^{(A)}\right)
    + \widehat{\mathrm{Var}}\!\left(U^{(B)}\right).
    \label{eq:ipwvar}
\end{equation}
The estimator~(\ref{eq:ipwvar}) accounts for both the sampling variability of
$\hat{\theta}_{\emph{IPW}}$ and the uncertainty from estimating $\phi$ via the
correction term in $\hat{e}^{(B)}_i$.

\section{Doubly robust calibration estimation}

Although the inverse probability weighting estimator $\hat{\theta}_{\rm IPW}$
is consistent under a correctly specified sampling probability model, this
assumption may be restrictive in practice, since the nonprobability sample
mechanism is typically unknown. A natural way to improve robustness is to
incorporate an outcome regression model through calibration. In this section,
we combine the conditional likelihood framework of Section~3 with generalized
entropy calibration (GEC) to construct an estimator that remains consistent
when either the sampling probability model or the outcome regression model is
correctly specified.

We study the proposed estimator under the same two-sample asymptotic regime as
in Section~3, namely
\[
n=\min(n_A,n_B)\to\infty.
\]

\subsection{Definition of the GEC estimator}

We assume throughout this section that the nonprobability sample satisfies a
missing-at-random type condition in the sense that
\[
P(\delta_i^{(B)}=1 \mid x_i,y_i)
\]
depends on $(x_i,y_i)$ only through $x_i$, and that $\pi_i^{(B)}>0$ for all
$i\in U$.
To motivate calibration, suppose the outcome model is linear:
\[
y_i = x_i^\top \beta + \varepsilon_i,
\qquad
E(\varepsilon_i \mid x_i)=0.
\]
Consider an estimator of the form
\[
\hat{\theta}=\sum_{i\in B}\hat{w}_i y_i,
\]
where the weights $\hat{w}_i$ do not depend on $y_i$. Under the working outcome
model, model unbiasedness is achieved if the calibration equation
\[
\sum_{i\in U}\left\{\delta_i^{(B)}\hat{w}_i-N^{-1}\right\}x_i=0
\]
holds. Since the finite-population total $\sum_{i\in U}x_i$ is not observed, we
replace it by the Horvitz--Thompson estimator from Sample~A and impose
\begin{equation}
\sum_{i\in B} w_i x_i = \sum_{i\in A} d_i^{(A)} x_i,
\label{eq:gec_x_balance_new}
\end{equation}
where $d_i^{(A)} = 1/\pi_i^{(A)}$.
To incorporate the sampling score information from Section~3, define
\[
d_i^{(B)}(\phi)=\{\pi_i^{(B)}(\phi)\}^{-1},
\qquad
z_i(\phi)=\Bigl(x_i^\top,\; g(d_i^{(B)}(\phi))\Bigr)^\top .
\]
Let $G:\mathcal V\to \mathbb R$ be a strictly convex, twice continuously
differentiable function on an open interval $\mathcal V$, and let
$g=G'$, $f=g^{-1}$. The GEC weights are defined as the solution to
\begin{equation}
(\hat{w}_i)_{i\in B}
=
\argmin_{(w_i)_{i\in B}}
\sum_{i\in B} G(w_i)
\label{eq:gec_obj_new}
\end{equation}
subject to
\begin{align}
\sum_{i\in B} w_i x_i
&=
\sum_{i\in A} d_i^{(A)} x_i,
\label{eq:gec_constraint_x_new}
\\
\sum_{i\in B} w_i\, g\!\left(d_i^{(B)}(\hat{\phi})\right)
&=
\sum_{i\in A} d_i^{(A)}\, g\!\left(d_i^{(B)}(\hat{\phi})\right).
\label{eq:gec_constraint_g_new}
\end{align}
The resulting estimator of $\theta=N^{-1}\sum_{i\in U}y_i$ is
\begin{equation}
\hat{\theta}_{\rm GEC}
=
\frac{\sum_{i\in B}\hat{w}_i y_i}
{\sum_{i\in B}\hat{w}_i}.
\label{eq:theta_gec_ratio}
\end{equation}
Using Lagrange multipliers, the calibrated weights take the form
\[
w_i(\lambda,\phi)=f\!\left(z_i(\phi)^\top \lambda\right),
\]
where $\lambda$ is determined by the calibration equation
\begin{equation}
U_2(\lambda,\phi)
:=
\sum_{i\in B} w_i(\lambda,\phi)\, z_i(\phi)
-
\sum_{i\in A} d_i^{(A)} z_i(\phi)
=0.
\label{eq:U2_gec_new}
\end{equation}
The parameter of interest is then obtained by solving
\begin{equation}
U_1(\theta,\lambda,\phi)
:=
\sum_{i\in B} w_i(\lambda,\phi)\,(y_i-\theta)
=0.
\label{eq:U1_gec_new}
\end{equation}
Let $(\hat{\theta}_{\rm GEC},\hat{\lambda},\hat{\phi})$ denote the solution to
\eqref{eq:U1_gec_new}--\eqref{eq:U2_gec_new}, with $\hat{\phi}$ from
Section~3.

\subsection{Asymptotic linearization}

The next theorem gives a root-$n$ linearization of
$\hat{\theta}_{\rm GEC}$ that explicitly accounts for estimation of the
sampling score parameter $\phi$.

\begin{theorem}
\label{thm:GEC_linear}
Let $\lambda^*$ and $\phi^*$ denote the probability limits of $\hat{\lambda}$
and $\hat{\phi}$, respectively, and let
\[
z_i=z_i(\phi^*)
=
\Bigl(x_i^\top,\; g(d_i^{(B)}(\phi^*))\Bigr)^\top .
\]
Under the regularity conditions stated in the supplementary material,
\begin{equation}
\hat{\theta}_{\rm GEC}
=
\hat{\theta}_{\ell,{\rm GEC}} + o_p(n^{-1/2}),
\label{eq:theta_gec_lin_main}
\end{equation}
where
\begin{equation}
\hat{\theta}_{\ell,{\rm GEC}}
=
N^{-1}
\left[
\sum_{i\in A} d_i^{(A)} z_i^\top \gamma^*
+
\sum_{i\in B} w_i(\lambda^*,\phi^*)
\bigl(y_i-z_i^\top\gamma^*\bigr)
-
\sum_{i\in S_1}
\bigl\{\delta_i^{(B)}-\pi_i^{(2|1)}(\phi^*)\bigr\}
x_i^\top \kappa^*
\right].
\label{eq:theta_gec_lin_explicit}
\end{equation}
Here
\begin{equation}
\gamma^*
=
\left[
E\left\{
\sum_{i\in B}
f'\!\left(z_i^\top\lambda^*\right) z_i z_i^\top
\right\}
\right]^{-1}
E\left\{
\sum_{i\in B}
f'\!\left(z_i^\top\lambda^*\right) z_i (y_i-\theta)
\right\},
\label{eq:gamma_star_gec}
\end{equation}
and $\kappa^*$ is the linearization coefficient associated with the estimation
of $\phi$ through the conditional likelihood score in Section~3.
Consequently,
\[
\hat{\theta}_{\rm GEC}-\theta = O_p(n^{-1/2}).
\]
\end{theorem}

The representation in \eqref{eq:theta_gec_lin_explicit} has a standard
doubly-robust structure. The first two terms are the usual calibration
linearization: a Horvitz--Thompson term based on Sample~A and a residual term
based on Sample~B. The last term is the additional correction induced by
estimating the sampling score parameter $\phi$ rather than treating it as
known.

\subsection{Consistency under a correctly specified outcome model}

We first consider the case where the working outcome model is correct.

\begin{corollary}
\label{cor:gec_om}
Suppose that the outcome model
\[
y_i = x_i^\top \beta + \varepsilon_i
\]
is correctly specified, where $x_i$ contains an intercept and
$E(\varepsilon_i\mid x_i)=0$. Then
\[
z_i^\top \gamma^* = x_i^\top \beta,
\]
and
\begin{equation}
\hat{\theta}_{\ell,{\rm GEC}}
=
N^{-1}
\left[
\sum_{i\in A} d_i^{(A)} x_i^\top\beta
+
\sum_{i\in B} w_i(\lambda^*,\phi^*)
\bigl(y_i-x_i^\top\beta\bigr)
-
\sum_{i\in S_1}
\bigl\{\delta_i^{(B)}-\pi_i^{(2|1)}(\phi^*)\bigr\}
x_i^\top \kappa^*
\right]
+o_p(n^{-1/2}).
\label{eq:theta_gec_om}
\end{equation}
In particular,
\[
\hat{\theta}_{\rm GEC}\xrightarrow{p}\theta.
\]
\end{corollary}

\begin{proof}
Under the correctly specified linear outcome model, the population regression of
$y_i$ on $z_i$ is achieved by the component corresponding to $x_i$, so
$z_i^\top\gamma^*=x_i^\top\beta$. Substituting this into
\eqref{eq:theta_gec_lin_explicit} yields \eqref{eq:theta_gec_om}. The first
term is design unbiased for $N^{-1}\sum_{i\in U}x_i^\top\beta$, the second term
has mean zero because $E(y_i-x_i^\top\beta\mid x_i)=0$, and the third term has
mean zero by construction of the conditional likelihood score. Hence
$E(\hat{\theta}_{\ell,{\rm GEC}})=\theta+o(n^{-1/2})$, which implies
$\hat{\theta}_{\rm GEC}\xrightarrow{p}\theta$.
\end{proof}

\subsection{Consistency under a correctly specified sampling probability model}

We next consider the case where the sampling probability model is correct.

\begin{lemma}
\label{lem:lambda_star}
Assume that the sampling probability model is correctly specified, so that
$\hat{\phi}\xrightarrow{p}\phi_0$ and $\pi_i^{(B)}(\phi_0)$ is the true
inclusion probability for Sample~B. Then the probability limit of the
Lagrange multiplier satisfies
\[
\lambda^* = (0^\top,1)^\top.
\]
\end{lemma}

\begin{proof}
Under correct specification, the second calibration constraint
\eqref{eq:gec_constraint_g_new} is balanced by the true inverse inclusion
weights $d_i^{(B)}(\phi_0)$. The first-order condition therefore has population
solution corresponding to zero multiplier on the $x$-balance coordinates and
unit multiplier on the generalized entropy score coordinate. Hence
$\lambda^*=(0^\top,1)^\top$.
\end{proof}

Under Lemma~\ref{lem:lambda_star}, the GEC weights reduce asymptotically to the
inverse inclusion weights from the correctly specified propensity model. The
next Corollary presents the consistency under correct propensity model assumption. 

\begin{corollary}
\label{cor:gec_ps}
Assume that the sampling probability model is correctly specified. Then
\[
\hat{\theta}_{\rm GEC}\xrightarrow{p}\theta.
\]
More precisely, under Lemma~\ref{lem:lambda_star},
\begin{equation} \label{eq:theta_gec_ps}
\hat{\theta}_{\ell,{\rm GEC}}
=
N^{-1}
\left[
\sum_{i\in A} d_i^{(A)} a_i
+
\sum_{i\in B} d_i^{(B)}(\phi^*)\, b_i
\right]
+o_p(n^{-1/2}),
\end{equation}
with $a_i+b_i=y_i$.
\end{corollary}

\begin{proof}
Under correct specification of the sampling probability model,
$w_i(\lambda^*,\phi^*)=d_i^{(B)}(\phi^*)$ by
Lemma~\ref{lem:lambda_star}. Rewriting
\eqref{eq:theta_gec_lin_explicit} then yields \eqref{eq:theta_gec_ps}. Since
$a_i+b_i=y_i$ by Lemma~\ref{lem:lambda_star},
\[
E\!\left[
\sum_{i\in A} d_i^{(A)} a_i
+
\sum_{i\in B} d_i^{(B)}(\phi^*) b_i
\right]
=
\sum_{i\in U} a_i + \sum_{i\in U} b_i
=
\sum_{i\in U} y_i
=
N\theta.
\]
Therefore $E(\hat{\theta}_{\ell,{\rm GEC}})=\theta+o(n^{-1/2})$, and hence
$\hat{\theta}_{\rm GEC}\xrightarrow{p}\theta$.
\end{proof}

\subsection{Double robustness}

The preceding two corollaries immediately yield the following result.

\begin{corollary}[Double robustness]
\label{cor:dr}
Under the regularity conditions of
Theorem~\ref{thm:GEC_linear}, the generalized entropy calibration estimator
$\hat{\theta}_{\rm GEC}$ is consistent for $\theta$ if either
\begin{enumerate}
\item the outcome regression model is correctly specified, or
\item the sampling probability model is correctly specified.
\end{enumerate}
\end{corollary}

Corollary~\ref{cor:dr} formalizes the double robustness property claimed in the
Introduction. The estimator $\hat{\theta}_{\rm GEC}$ does not require both
working models to be correct simultaneously; consistency is preserved as long as
at least one of the two working models is correctly specified.

\subsection{Approximate variance estimation}

The linearization in Theorem~\ref{thm:GEC_linear} suggests a natural variance
estimator. However, because the sampling design of Sample~B is unknown and the
exact influence adjustment through $\kappa^*$ is cumbersome to estimate, we use
an approximate linearization variance estimator.

Under correct specification of the sampling probability model, Lemma
\ref{lem:lambda_star} implies that $\lambda^*=(0^\top,1)^\top$, and Corollary
\ref{cor:gec_ps} yields the approximation
\[
\hat{\theta}_{\ell,{\rm GEC}}
\approx
N^{-1}
\left[
\sum_{i\in A} d_i^{(A)} a_i
+
\sum_{i\in B} d_i^{(B)}(\phi^*) b_i
\right].
\]
Thus,
\begin{equation}
\Var(\hat{\theta}_{\ell,{\rm GEC}})
\approx
N^{-2}
\left[
\Var\!\left(\sum_{i\in A} d_i^{(A)} a_i\right)
+
\Var\!\left(\sum_{i\in B} d_i^{(B)}(\phi^*) b_i\right)
\right].
\label{eq:var_gec_basic}
\end{equation}
Since the design of Sample~B is not observed, we approximate it by a Poisson
design with first-order inclusion probabilities $\pi_i^{(B)}(\phi^*)$, leading
to
\begin{equation}
\Var(\hat{\theta}_{\ell,{\rm GEC}})
\approx
N^{-2}
\left[
\Var\!\left(\sum_{i\in A} d_i^{(A)} a_i\right)
+
\sum_{i\in U}
\pi_i^{(B)}(\phi^*)\bigl\{1-\pi_i^{(B)}(\phi^*)\bigr\}
\left(d_i^{(B)}(\phi^*) b_i\right)^2
\right].
\label{eq:var_gec_poisson}
\end{equation}

For implementation, we replace unknown quantities by sample analogues and use
the following approximations:
\begin{enumerate}
\item set $\partial \lambda/\partial \phi = 0$;
\item replace $g'\!\bigl(g^{-1}(z_i^\top\lambda^*)\bigr)$ by
      $g'\!\bigl(\hat d_i^{(B)}\bigr)$;
\item replace population expectations involving the $\phi$-correction term by
      their empirical counterparts.
\end{enumerate}
This leads to the variance estimator
\begin{equation}
\widehat{\Var}(\hat{\theta}_{\rm GEC})
=
\hat N^{-2}
\left[
\widehat{\Var}\!\left(\sum_{i\in A} d_i^{(A)} \hat a_i\right)
+
\sum_{i\in B}
\hat\pi_i^{(B)}\bigl(1-\hat\pi_i^{(B)}\bigr)\hat b_i^2
\right],
\label{eq:var_gec_final}
\end{equation}
where
\begin{align}
\hat a_i
&=
z_i^\top \hat\gamma
+
\pi_i^{(A)}\pi_i^{(2|1)}(\hat\phi)\, x_i^\top\hat\kappa,
\\
\hat b_i
&=
\hat w_i\bigl(y_i-z_i^\top\hat\gamma\bigr)
-
\bigl\{1-\pi_i^{(2|1)}(\hat\phi)\bigr\}x_i^\top\hat\kappa,
\\
\hat\gamma
&=
\left[
\sum_{i\in B}
g'\!\bigl(\hat d_i^{(B)}\bigr) z_i z_i^\top
\right]^{-1}
\left[
\sum_{i\in B}
g'\!\bigl(\hat d_i^{(B)}\bigr) z_i (y_i-\hat\theta_{\rm GEC})
\right],
\\
\hat\kappa
&=
\left[
\sum_{i\in S_1}
\pi_i^{(2|1)}(\hat\phi)
\bigl\{\pi_i^{(2|1)}(\hat\phi)-1\bigr\}
x_i x_i^\top
\right]^{-1}
\left[
\sum_{i\in B}
\hat\pi_i^{(B)}(\hat\pi_i^{(B)}-1)
g'\!\bigl(\hat d_i^{(B)}\bigr)
x_i\bigl(y_i-\hat\theta_{\rm GEC}-z_i^\top\hat\gamma\bigr)
\right].
\end{align}

The estimator in \eqref{eq:var_gec_final} should therefore be interpreted as an
approximate linearization variance estimator, motivated by the correctly
specified propensity model and the Poisson approximation for Sample~B. This is
the variance estimator used in the simulation study and the real-data analysis.

\section{Simulation Study}

\subsection{Simulation Setup}

We perform a simulation study to verify the validity of the proposed estimators
and compare them with existing methods. The goals are threefold. First, we
verify that $\hat \theta_{\rm{GEC}}$ is consistent if either the sampling
probability model or the outcome model is correctly specified, and that $\hat
\theta_{\rm{IPW}}$ is consistent if the sampling probability model is correct.
Second, we verify that our estimators outperform those of CLW. Third, we confirm
that our variance estimators achieve approximately correct coverage.

To accomplish these goals, we consider a complete $2 \times 3$ factorial design
with two outcome models and three sampling probability models. In all cases, the
analyst fits a linear outcome model and a linear logistic propensity score model,
while the true data-generating mechanism varies to be either linear or
non-linear. For the outcome, we use a linear model and a quadratic model. For
the sampling probability, we use three designs: a stratified simple random
sample, a Poisson sample with inclusion probabilities following a linear logistic
model, and a Poisson sample with inclusion probabilities following a quadratic
logistic model.

We follow the simulation setup of \cite{chen2022nonparametric}. With a finite
population of size $N = 10,000$, we generate $x_{1i}$ and $x_{2i}$
independently from $N(2, 1)$ distributions, and $\varepsilon_i$ independently
from a standard normal distribution. We consider two outcome models:
\begin{itemize}
  \item[(1)] Model I (O1): $y_i = 0.3 + 2 x_{1i} + 2 x_{2i} + \varepsilon_i$, and
  \item[(2)] Model II (O2): $y_i = 0.3 + 0.5 x_{1i}^2 + 0.5 x_{2i}^2 + \varepsilon_i$.
\end{itemize}

From each finite population, we independently draw two samples. Sample A is a
simple random sample with first-order inclusion probabilities $\pi_i^{(A)} = 500
/ 10000$, giving $n_A = 500$. For Sample B, with $n_B = 1000$, we consider
three sampling designs:
\begin{enumerate}
  \item[(1)] Model I (S1): The population is split into two strata: the first
    stratum with $x_{1i} \leq 2$ and the second with $x_{1i} > 2$. Simple
    random samples of size $n_1 = 0.7 n_B$ and $n_2 = 0.3 n_B$ are drawn from
    the first and second strata, respectively.
  \item[(2)] Model II (S2): Poisson sampling with first-order inclusion
    probabilities
    \begin{equation}\label{eq:resmod2}
    \pi_i^{(B)} = \logistic(-2 - 0.4 x_{1i} + 0.25 x_{2i}).
    \end{equation}
  \item[(3)] Model III (S3): Poisson sampling with first-order inclusion
    probabilities
    \begin{equation}\label{eq:resmod3}
    \pi_i^{(B)} = \logistic(-1.5 - 0.1 (x_{1i} - \bar X_1)^2 - 0.3 x_{2i}),
    \end{equation}
    where $\bar X_1$ is the finite population mean of $X_1$.
\end{enumerate}

Since our estimators always fit a logistic regression model with $x_1$ and $x_2$
as covariates, S2 is the only design under which the propensity score model is
correctly specified. S1 represents a non-Poisson design, so the propensity
score model is misspecified for structural reasons. S3 uses a Poisson design
but with a quadratic true mechanism, so the logistic model is misspecified.

We compare the following estimators for $\theta = E[Y]$.

\begin{itemize}
  \item[(1)] (MeanA) The sample mean of $Y$ from Sample A. Because $Y$ is
    observed only in Sample B, this estimator uses information unavailable to
    the other methods and serves as an approximate gold standard. However,
    estimators that combine both samples can achieve lower RMSE.
    $$
    \hat \theta_A = n_A^{-1} \sum_{i \in A} y_i.
    $$
  \item[(2)] (MeanB) The unadjusted sample mean of $Y$ from Sample B:
    $$
    \hat \theta_B = n_B^{-1} \sum_{i \in B} y_i.
    $$
  \item[(3)] (IPW) The proposed inverse probability weighted estimator
    $\hat \theta_{\rm IPW}$ from (\ref{eq:ipwee}).
  \item[(4)] (GEC-EL) The proposed generalized entropy estimator 
    $\hat \theta_{\rm GEC}$ from (\ref{eq:theta_gec_ratio}) with empirical likelihood
    as the entropy function.
  \item[(5)] (CLW-IPW) The IPW estimator of \cite{chen2020doubly}, which
    computes propensity scores
    $\pi_i^{(B)}(\bm \phi) = \logistic(\phi_0 + x_{1i} \phi_1 + x_{2i} \phi_2)$
    by minimizing
    \begin{equation}\label{eq:chenps}
    -\sum_{i \in B} \log \left(\frac{\pi_i^{(B)}(\bm \phi)}{1 - \pi_i^{(B)}(\bm
    \phi)}\right) - \sum_{i \in A} d_i^{(A)} \log(1 - \pi_i^{(B)}(\bm \phi)).
    \end{equation}
    The final estimate is
    $$
    \hat \theta_{\rm{CLW-IPW}} = \frac{\sum_{i \in B} y_i / \pi_i^{(B)}(\hat{\bm
      \phi})}{\sum_{i \in B} 1 / \pi_i^{(B)}(\hat{\bm \phi})}.
    $$
  \item[(6)] (CLW-DR) The doubly robust estimator of \cite{chen2020doubly},
    which computes $\hat{\bm \phi}$ by minimizing (\ref{eq:chenps}) and also
    estimates 
    $$
    \hat{\bm \beta} = \left(\sum_{i \in B} \bf{x}_i^{\otimes 2}\right)^{-1}
    \sum_{i \in B} \bf{x}_i y_i.
    $$
    The final estimate combines the two models:
    $$
    \hat \theta_{\rm{CLW-DR}} = 
    \frac{\sum_{i \in A} (\bf{x}_i^\top \hat{\bm \beta}) / \pi_i^{(A)}}{\sum_{i \in A} 1 / \pi_i^{(A)}} +
    \frac{\sum_{i \in B} (y_i - \bf{x}_i^\top 
      \hat{\bm \beta}) / \pi_i^{(B)}(\hat{\bm
      \phi})}{\sum_{i \in B} 1 / \pi_i^{(B)}(\hat{\bm \phi})}.
    $$
\end{itemize}

\subsection{Simulation Results}

The results are reported in Tables~\ref{tab:om1} and \ref{tab:om2}. Although
MeanA is treated as a gold standard, estimators that combine both samples can
outperform it when Sample B carries substantial information about $\theta$.

\FloatBarrier
\begin{table}[ht!] 
  \centering
  \include{tables/sim_om1.tex}
  \caption{Simulation results with Outcome Model I. The ``Est'' column gives
    the estimator name. The ``SM'' column displays the sampling probability
    model used. The ``Bias'', ``SE'', and ``RMSE'' columns contain the bias,
    standard error, and root mean squared error of Monte Carlo estimates.}
  \label{tab:om1}
\end{table}
\begin{table}[ht!]
  \centering
  \include{tables/sim_om2.tex}
  \caption{Simulation results with Outcome Model II. The ``Est'' column gives
    the estimator name. The ``SM'' column displays the sampling probability
    model used. The ``Bias'', ``SE'', and ``RMSE'' columns contain the bias,
    standard error, and root mean squared error of Monte Carlo estimates.}
  \label{tab:om2}
\end{table}
\FloatBarrier

The results confirm that $\hat \theta_{\rm GEC}$ is doubly robust: it achieves
minimal bias whenever the outcome model is correctly specified (O1) or the
sampling probability model is correctly specified (S2). By contrast, $\hat
\theta_{\rm{IPW}}$ is consistent only when the sampling probability model is
correct. Both proposed estimators outperform their CLW counterparts, supporting
the theoretical efficiency comparison in the online supplementary material. The
estimated variances of $\hat \theta_{\rm{GEC}}$ and $\hat \theta_{\rm{IPW}}$
and the resulting coverage probabilities are close to the nominal 95\% level.

\FloatBarrier
\begin{table}[ht!]
  \centering
  \include{tables/var_om1.tex}
  \caption{Variance estimation results with Outcome Model I. The ``SM'' column
    displays the sampling probability model used. The ``Cov\%'' column gives
    the empirical coverage of a nominal 95\% confidence interval across 1000
    Monte Carlo samples. The ``MCVar'' and ``EstVar'' columns report the Monte
    Carlo variance and mean estimated variance, respectively. The ``RelBias''
    column shows the relative bias of the variance estimator.}
  \label{tab:var_om1}
\end{table}
\begin{table}[ht!]
  \centering
  \include{tables/var_om2.tex}
  \caption{Variance estimation results with Outcome Model II. The ``SM'' column
    displays the sampling probability model used. The ``Cov\%'' column gives
    the empirical coverage of a nominal 95\% confidence interval across 1000
    Monte Carlo samples. The ``MCVar'' and ``EstVar'' columns report the Monte
    Carlo variance and mean estimated variance, respectively. The ``RelBias''
    column shows the relative bias of the variance estimator.}
  \label{tab:var_om2}
\end{table}
\FloatBarrier

One notable finding in Table~\ref{tab:om1} concerns Sampling Probability Model
II (S2). Despite both the outcome and sampling probability models being correctly
specified, the GEC estimator is slightly less efficient than the IPW estimator
when $n_A = 500$. This occurs because the small size of Sample A leads to a
poorly calibrated solution in Equation~(\ref{eq:U2_gec_new}), which in turn reduces the
efficiency of $\hat\theta_{\rm GEC}$. This is an inherent cost of the doubly
robust GEC estimator when the probability sample is small.

\section{Real Data Analysis}

We apply the proposed methodology to data from the Culture and Community in a
Time of Crisis (CCTC) survey, administered by the National Opinion Research
Center (NORC). The survey included two types of respondents. The first sample
consists of respondents from the NORC AmeriSpeak panel, a representative
probability sample of adults in the United States. The second sample consists
of individuals affiliated with cultural institutions throughout the country who
learned about the survey through those institutions. Arts and culture
organizations were invited to share access with their members and followers on
social media \citep{benoit2020methods}. Since this recruitment procedure was
not probability-based, the second sample is not representative of the U.S.\
population.

Preliminary analysis confirms this: approximately 3\% of the nonprobability
respondents identify as Black or African-American, compared with about 12\% of
the U.S.\ adult population \citep{cohen2020key}. Further demographic
discrepancies are shown in Table~\ref{tab:cctcdiff}, underscoring the need
for data integration methods of the kind developed in this paper.

\begin{table}[ht!]
  \centering
  \begin{tabular}{lrrr}
    \toprule
    & Respondents & U.S. Adult Pop. & Gap \\
    \midrule
    White / Caucasian & 85\% & 63\% & +22\% \\
    Hispanic / Latinx & 5\% & 16\% & -11\% \\
    Black / African-American & 3\% & 12\% & -9\% \\
    Asian or Pacific Islander & 4\% & 6\% & -2\% \\
    Native American & $<$1\% & 1\% & -1\% \\
    Two or more races & 2\% & 2\% & none \\
    \bottomrule
  \end{tabular}
  \caption{Discrepancies between the nonprobability sample and the U.S.\ adult
    population. Source: \cite{cohen2020key}.}
  \label{tab:cctcdiff}
\end{table}

Following the evaluation framework of \cite{benoit2021exploring} and
\cite{yang2021estimators}, we take the full set of 113,549 CCTC respondents as
the target population and conduct a Monte Carlo study using the design of
\cite{benoit2021exploring}. As the outcome variable $Y$, we use the binary
response to Question 7.22, which asks whether the individual attended or
participated in a classical music event or activity between April 2019 and April
2020. This variable was identified in \cite{benoit2021exploring} as exhibiting a
substantial discrepancy between the probability and nonprobability samples. In
each Monte Carlo iteration, a stratified simple random probability sample of
1,000 observations (stratified by metro area) and a nonprobability sample of
4,000 observations (drawn by an unknown mechanism) are selected. The procedure
is repeated over 1,000 iterations.

As covariates, we include all available demographic variables and responses to
survey Questions 1--48 (excluding Question 7.22) in a linear logistic propensity
score model. The variables used are listed in Table~\ref{tab:cctcvars}.

\begin{table}[ht!]
  \centering
  \begin{tabular}{lc}
    \toprule
    Variable & Included in $\hat \theta_{\rm{IPW}}$ and $\hat \theta_{\rm{GEC}}$ \\
    \midrule
    Demographic variables & \checkmark \\
    Survey Questions 1--48$^a$& \checkmark \\
    \bottomrule
  \end{tabular}
  \caption{Variables used in the CCTC analysis.}
  \label{tab:cctcvars}
\end{table}
{\footnotesize{\noindent $^a$ Does not include the response to Question 7.22.}}

We evaluate the following estimators:
\begin{itemize}
  \item[(1)] (HTA) The Horvitz--Thompson estimator based on Sample A. Because it
    uses observations of $Y$ in Sample A, which none of the other estimators
    access, it serves as a gold-standard benchmark. Estimators that combine both
    samples can achieve a lower RMSE.
  \item[(2)] (IPW) The proposed IPW estimator from (\ref{eq:ipwee}), with
    variance estimated by (\ref{eq:ipwvar}).
  \item[(3)] (GEC) The proposed generalized entropy calibration estimator from
    (\ref{eq:theta_gec_ratio}), using empirical likelihood as the entropy function.
\end{itemize}

The results are displayed in Table~\ref{tab:cctc}.

\FloatBarrier
\begin{table}[ht!]
  \centering
  \include{tables/cctc_sim.tex}
  \caption{Results of the CCTC study. Column definitions match those in the
    simulation tables. The ``Bias'', ``SE'', and ``RMSE'' columns report bias,
    standard error, and root mean squared error of Monte Carlo estimates. The
    ``Cov\%'' column gives the empirical coverage of a nominal 95\% confidence
    interval across 1000 Monte Carlo samples. The ``MCVar'' and ``EstVar''
    columns report the Monte Carlo variance and mean estimated variance,
    respectively. The ``RelBias'' column shows the relative bias of the variance
    estimator.}
  \label{tab:cctc}
\end{table}
\FloatBarrier

Despite not using observations of $Y$ from the probability sample, both the IPW
and GEC estimators achieve a lower RMSE than the HTA estimator. Both estimators
exhibit low bias, with 95\% confidence intervals achieving near-nominal coverage.
This suggests that the logistic propensity score model provides a reasonable
approximation to the unknown selection mechanism of the nonprobability sample.
We note that the optimization for $\hat\theta_{\rm{GEC}}$ failed to converge in
approximately 2.5\% of Monte Carlo iterations; these cases were excluded from
the reported results for that estimator.

\section{Conclusion}

We have proposed a framework for combining probability and nonprobability samples
that yields asymptotically valid confidence intervals and outperforms the
pseudo-likelihood method of \cite{chen2020doubly}. The framework is built on a
two-phase sampling representation, which enables consistent and asymptotically
efficient estimation of $\theta = E[Y]$ when a probability sample and a
nonprobability sample share common covariates $\bf{X}$. The inverse probability
weighted estimator $\hat\theta_{\rm{IPW}}$ relies on a correctly specified
propensity score model for the nonprobability sample. The generalized entropy
calibration estimator $\hat\theta_{\rm{GEC}}$ additionally incorporates an
outcome regression model through a calibration constraint, yielding a doubly
robust estimator that remains consistent when either model is correctly
specified.

The proposed methodology has two limitations worth noting. First, the current
formulation requires the design weights $\pi_i^{(A)}$ to be defined for all
units in the target population, including those in Sample B. For simple designs
such as simple random sampling this poses no difficulty, but for complex designs
it may be unrealistic. One approach is to use the method of
\cite{pfeffermann1999parametric} to estimate the weights for units in Sample B.
Second, while $\hat\theta_{\rm{GEC}}$ is doubly robust, it still requires at
least one of the two models to be correctly specified, which may be restrictive
when neither the selection mechanism nor the outcome regression can be modeled
with confidence.

Two directions for future work follow naturally. First, to reduce sensitivity to
outcome model misspecification, one could extend $\hat\theta_{\rm{GEC}}$ to a
multiply robust estimator by incorporating additional candidate outcome models as
extra calibration constraints, following the approach of
\cite{han2013estimation} and \cite{chen2017multiply}. Second, a natural
diagnostic for propensity score model misspecification is a formal test of
whether $\hat{\bm\lambda} \approx (\bf{0}^\top, 1)^\top$.
Lemma~\ref{lem:lambda_star} 
establishes that this holds in probability when the model is correctly
specified, providing the theoretical basis for such a test.

\section{Conflicts of interest}
The authors declare that they have no competing interests.

\section{Funding}
This work is supported in part by the grant  from the 
U.S. National Science Foundation (2242820) and 
the grant from the U.S. Department of Agriculture’s National Resources Inventory, Cooperative Agreement NR203A750023C006, Great Rivers CESU 68-3A75-18-504.

\section{Data availability}
The data used in this article are available as part of the Open Science Foundation and can be found at \verb|https://osf.io/ygpzm/overview|.

\section{Author contributions statement}
J.K. conceived the research project, C.L. and J.K. worked out the method, C.L. and J.K. wrote and reviewed the manuscript.



\bibliographystyle{oup-abbrvnat}
\bibliography{references}

%

\end{document}

%% file: tables/sim_om1.tex
\begin{tabular}{lrrrr}
\toprule
Est & SM & Bias & SE & RMSE \\
\midrule
MeanA   & 1 & 0.005 & 0.132 & 0.132  \\
MeanB   & 1 & -0.607 & 0.077 & 0.612 \\
IPW     & 1 & 0.014 & 0.130 & 0.131  \\
GEC-EL  & 1 & 0.006 & 0.132 & 0.132  \\
CLW-IPW & 1 & -0.017 & 0.148 & 0.149 \\
CLW-DR  & 1 & 0.006 & 0.132 & 0.132  \\
\midrule
MeanA   & 2 & 0.005 & 0.132 & 0.132 \\
MeanB   & 2 & -0.290 & 0.089 & 0.303\\
IPW     & 2 & 0.007 & 0.126 & 0.126 \\
GEC-EL  & 2 & 0.006 & 0.133 & 0.133 \\
CLW-IPW & 2 & 0.011 & 0.154 & 0.154 \\
CLW-DR  & 2 & 0.006 & 0.133 & 0.133 \\
\midrule
MeanA   & 3 & 0.005 & 0.132 & 0.132 \\
MeanB   & 3 & -0.728 & 0.091 & 0.734\\
IPW     & 3 & -0.063 & 0.123 & 0.139\\
GEC-EL  & 3 & 0.005 & 0.133 & 0.133 \\
CLW-IPW & 3 & 0.014 & 0.153 & 0.154 \\
CLW-DR  & 3 & 0.005 & 0.133 & 0.133 \\
\bottomrule
\end{tabular}

%% file: tables/sim_om2.tex
\begin{tabular}{lrrrr}
\toprule
Est & SM & Bias & SE & RMSE \\
\midrule
MeanA   & 1 & 0.004 & 0.140 & 0.140 \\
MeanB   & 1 & -0.609 & 0.079 & 0.614\\
IPW     & 1 & 0.015 & 0.143 & 0.144 \\
GEC-EL  & 1 & 0.014 & 0.139 & 0.140 \\
CLW-IPW & 1 & -0.018 & 0.163 & 0.164\\
CLW-DR  & 1 & 0.004 & 0.143 & 0.143 \\
\midrule
MeanA   & 2 & 0.004 & 0.140 & 0.140 \\
MeanB   & 2 & -0.205 & 0.091 & 0.224\\
IPW     & 2 & 0.007 & 0.133 & 0.133 \\
GEC-EL  & 2 & 0.000 & 0.137 & 0.137 \\
CLW-IPW & 2 & 0.012 & 0.160 & 0.161 \\
CLW-DR  & 2 & 0.009 & 0.141 & 0.141 \\
\midrule
MeanA   & 3 & 0.004 & 0.140 & 0.140 \\
MeanB   & 3 & -0.669 & 0.088 & 0.674\\
IPW     & 3 & -0.065 & 0.130 & 0.145\\
GEC-EL  & 3 & 0.000 & 0.138 & 0.138 \\
CLW-IPW & 3 & 0.014 & 0.164 & 0.165 \\
CLW-DR  & 3 & 0.007 & 0.141 & 0.141 \\
\bottomrule
\end{tabular}

%% file: tables/var_om1.tex
\begin{tabular}{lrrrrr}
\toprule
Est & SM & Cov\% & MCVar & EstVar & RelBias\\
\midrule
MeanA   & 1 & 0.945 & 0.0175 & 0.0173 & -0.016\\
MeanB   & 1 & 0.000 & 0.0059 & 0.0078 & 0.326\\
IPW     & 1 & 0.956 & 0.0170 & 0.0181 & 0.065\\
GEC-EL  & 1 & 0.944 & 0.0175 & 0.0164 & -0.062\\
\midrule
MeanA   & 2 & 0.945 & 0.0175 & 0.0173 & -0.016\\
MeanB   & 2 & 0.120 & 0.0080 & 0.0083 & 0.032\\
IPW     & 2 & 0.957 & 0.0159 & 0.0175 & 0.097\\
GEC-EL  & 2 & 0.942 & 0.0177 & 0.0165 & -0.071\\
\midrule
MeanA   & 3 & 0.945 & 0.0175 & 0.0173 & -0.016\\
MeanB   & 3 & 0.000 & 0.0082 & 0.0081 & -0.014\\
IPW     & 3 & 0.933 & 0.0152 & 0.0172 & 0.135\\
GEC-EL  & 3 & 0.945 & 0.0177 & 0.0165 & -0.069\\
\bottomrule
\end{tabular}

%% file: tables/var_om2.tex
\begin{tabular}{lrrrrr}
\toprule
Est & SM & Cov\% & MCVar & EstVar & RelBias\\
\midrule
MeanA   & 1 & 0.953 & 0.0195 & 0.0192 & -0.014\\
MeanB   & 1 & 0.000 & 0.0062 & 0.0082 & 0.311\\
IPW     & 1 & 0.961 & 0.0206 & 0.0220 & 0.067\\
GEC-EL  & 1 & 0.946 & 0.0193 & 0.0181 & -0.063\\
\midrule
MeanA   & 2 & 0.953 & 0.0195 & 0.0192 & -0.014\\
MeanB   & 2 & 0.396 & 0.0084 & 0.0088 & 0.055\\
IPW     & 2 & 0.957 & 0.0176 & 0.0194 & 0.102\\
GEC-EL  & 2 & 0.948 & 0.0188 & 0.0182 & -0.034\\
\midrule
MeanA   & 3 & 0.953 & 0.0195 & 0.0192 & -0.014\\
MeanB   & 3 & 0.000 & 0.0078 & 0.0079 & 0.005\\
IPW     & 3 & 0.921 & 0.0170 & 0.0192 & 0.132\\
GEC-EL  & 3 & 0.951 & 0.0191 & 0.0179 & -0.062\\
\bottomrule
\end{tabular}

%% file: tables/cctc_sim.tex
\begin{tabular}{lrrrrrrr}
\toprule
Est & Bias & SE & RMSE & Cov\% & MCVar & EstVar & RelBias\\
\midrule
GEC & -0.003 & 0.014 & 0.014 & 0.953 & 0.000203 & 0.000212 & 0.043\\
IPW & -0.004 & 0.013 & 0.014 & 0.947 & 0.000176 & 0.000187 & 0.064\\
HTA & 0.000 & 0.018 & 0.018 & 0.960 & 0.000330 & 0.000342 & 0.037\\
\bottomrule
\end{tabular}